\documentclass[%
 aip,
 amsmath,amssymb,
 reprint,%
]{revtex4-1}

\usepackage{graphicx}
\usepackage{dcolumn}
\usepackage{bm}

\usepackage[utf8]{inputenc}
\usepackage[T1]{fontenc}
\usepackage{mathptmx}
\usepackage{etoolbox}

\usepackage{gnuplot-lua-tikz}

\usepackage{amssymb}
\usepackage{amsthm}
\usepackage{amsmath}
\usepackage{dsfont}
\usepackage{xcolor}
\usepackage{booktabs} 
\usepackage{graphicx}

\DeclareMathOperator{\Tr}{Tr}
\newcommand{\Ell}{\mathcal{L}}
\newcommand{\R}{\mathds{R}}
\newcommand{\C}{\mathds{C}}
\newcommand{\bfa}{\boldsymbol{\alpha}}

\newcommand{\psib}{\boldsymbol{\psi}}
\newcommand{\eb}{{\mathbf e}}

\makeatletter
\def\@email#1#2{%
 \endgroup
 \patchcmd{\titleblock@produce}
  {\frontmatter@RRAPformat}
  {\frontmatter@RRAPformat{\produce@RRAP{*#1\href{mailto:#2}{#2}}}\frontmatter@RRAPformat}
  {}{}
}%
\makeatother
\begin{document}

\preprint{AIP/123-QED}

\title[]{{Quantum Optimal Control for Pure-State Preparation Using One Initial State}}

\author{Stefanie G{\"u}nther}
 \email{guenther5@llnl.gov}
  \affiliation{Lawrence Livermore National Laboratory, Livermore CA 94550}
\author{N. Anders Petersson}
  \affiliation{Lawrence Livermore National Laboratory, Livermore CA 94550}
\author{Jonathan L. DuBois}
  \affiliation{Lawrence Livermore National Laboratory, Livermore CA 94550}

\date{\today}

\begin{abstract}

This paper presents a  {framework for solving the pure-state preparation problem using numerical optimal control. As an example, we consider the case where} a number of qubits are dispersively coupled to a readout cavity. We model open system quantum dynamics using {the Markovian} Lindblad master equation, driven by external control pulses.
{The main result of this paper develops} a basis of density matrices (a parameterization) where each basis element is a density matrix itself. 
{Utilizing a specific objective function,} we show how an {ensemble} of the basis elements can be used {as a single
initial state throughout the optimization process} - independent of the system dimension. 
{We apply the general framework to the specific application of} ground-state reset of one and two qubits coupled to a readout cavity.
\end{abstract}

\maketitle

\section{Introduction}\label{sec:intro}

A key requirement for building general purpose quantum computers is the ability to {perform state preparation, i.e.~to initialize the qubits into a well defined state, such as the ground state}\cite{divincenzo2000physical}. 
When quantum algorithms are executed many times (to gather statistics of the results, or for variational algorithms that are restarted many times with modified parameters) initialization of the qubits can become a major bottleneck if its duration is long compared to the gate operations\cite{egger2018pulsed, peruzzo2014variational}. Further, {most schemes for fault tolerant quantum computation} require a continuous supply of qubits in a low-entropy state such as the ground state,~\cite{divincenzo2000physical, barrett2013simulating, reed2012realization} {magic states\cite{PhysRevA.71.022316}, or graph states\cite{PhysRevA.69.062311}} -- at short timescales and with high fidelity.

{In general, unconditional state preparation requires the system under study to be coupled in some way to a quantum dissipative channel\cite{weiss2012quantum} allowing for entropy flow from the system of interest to a bath. As such, engineering these dissipative processes through a combination of device design and driven control represents a fundamental challenge, and concomitant opportunity, for realization of useful quantum information processing systems (see, e.g. Refs.~\onlinecite{kapit2017upside,murch2012cavity,poyatos1996quantum}).}
In recent years, and with increasing computational power of classical computers, numerical optimal control has attracted much attention for {shaping the control pulses to} drive a quantum system to a desired target state (see e.g. the reviews in Refs.~\onlinecite{glaser2015training, koch2016controlling}, and references therein). While most prominently applied to shape pulses that realize logical gate operations~\cite{doi:10.1080/09500340802344933, PhysRevLett.89.188301}, {optimal control has also been applied for the preparation of quantum states, such as the ground state}~\cite{basilewitsch2019reservoir, boutin2017resonator,fischer2019time, abdelhafez2019gradient,basilewitsch2021fundamental}.

{Optimal control methods determine a set of control waveforms by minimizing an objective function, $J$, that represents the mismatch between the realized state and the target state at a final time.}
{In this paper, we present a numerical optimal control framework for shaping control pulses to drive open quantum systems to a pure target state.}
{We consider unconditional state preparation, where the initial state of the system under consideration is unknown, such that the control pulses must drive \textit{any} initial state to the common target. Therefore, the minimization has to take multiple initial states into account:}
{
\begin{align}
\min \frac 1M \sum_{i=1}^M J\left(\rho_m(T)\right),
\end{align}
where $\rho_m(T)$ represents the quantum state at the final time $T$, corresponding to the $m$-th initial state $\rho_m(0)$. In general, the $M$ initial states must span a basis for all possible initial states at time $t=0$, such that the objective function needs to be
evaluated for $M=N^2$ initial basis states, with $N$ being the underlying Hilbert space dimension.
Since the computational complexity for evolving one initial state to the final time $T$ scales with $N^2$, and} $N$ itself scales exponentially with the number of qubits, this quickly becomes computationally intractable. {As a result, optimal control applications for quantum state preparation are often limited} to small quantum systems{, or considering sub-spaces} in which a basis can be taken into account, and/or considering only a few specific (e.g. pure) initial qubit and/or cavity states\cite{basilewitsch2019reservoir, basilewitsch2019quantum, boutin2017resonator, fischer2019time, basilewitsch2017beating}.

{In this paper, we develop a basis of density matrices in such a way that only {\em one} initial condition needs to be taken into account during the optimization, $M=1$, independent of the Hilbert space dimension $N$. The proposed basis consists of $N^2$ density matrices that span all initial states in the Hilbert space. Any state can therefore be parameterized in this basis.}
Utilizing a specifically designed objective function {that is linear in the initial condition} then allows us to employ {an ensemble}
of the basis states as the only initial state that needs to be taken into account during the optimization. 
{Never the less,} the optimized control pulses drive \textit{any} initial quantum state to the {desired} pure state, hence achieving {optimal control} for pure-state preparation, with drastically reduced computational complexity. 
{As a result, optimal control for pure-state preparation becomes feasible for larger quantum systems.}

When applying {quantum optimal control} for unitary gate optimization, {where the target state is a unitary transformation of the initial state (often for realizing a logical operation), a similar result has been derived in Ref.~\onlinecite{Reich-Gualdi-Koch-2013, goerz2014optimal}, showing that it is sufficient to consider only} three specific initial states during the optimization, {M=3}, independent of the Hilbert space dimension. The three initial states are derived in such a way that they can distinguish between any unitary transformations within the Hilbert space. 
{For a unitary gate optimization, the objective function must take both the initial and the final states into account, e.g.~using the trace distance or the Hilbert-Schmidt projection between the unitarily transformed initial and the realized states. In this case, both states depend on the initial state and as a result, the objective function becomes nonlinear in the initial condition. In contrast, unconditional pure-state preparation considers a common target state that is independent of the initial condition, such that the objective function can be constructed to depend only linearly on the 
initial condition, allowing us to utilize an ensemble of all basis states as the only initial condition during the optimization procedure.}

{
The paper is structured as follows: Section \ref{sec:optim} presents the new density matrix parameterization as well as their ensemble state and the objective function for the proposed optimal control framework for pure-state preparation.} We then introduce the governing equations of the underlying open quantum dynamics in terms of Lindblad's master equation in Section \ref{sec:governingeq}. 
While the {proposed} basis of density matrices and objective function, together with the {single} initial condition, can be readily utilized in popular {optimal quantum control algorithms}, such as the Grape~\cite{khaneja2005optimal, schulte2011optimal} and the Krotov~\cite{sklarz2002loading,reich2012monotonically} algorithms, we present and employ an alternative optimal control strategy in Section \ref{sec:numericalapproach}, as implemented in the open source software Quandary \cite{quandaryGithub}.
{As a demonstration of the approach, Section \ref{sec:numerical_results} then presents numerical results for driving one and two qudits coupled to a cavity to the ground state of a coupled qudit-(qudit-)cavity system.}
{Further details on the proposed basis of density matrices are presented in Appendix \ref{app:basismats_proof} and \ref{app:N2basismats}. Appendix \ref{app:general_pure_state} and \ref{app:results_purestateprep} generalize the approach to the preparation of other pure states, besides the ground state.}

\section{One Initial Condition for Pure-State Optimization}\label{sec:optim}

{The pure-state preparation problem}
aims to find a set of control pulses 
that {drive any initial state at $t=0$ to a desired pure target state at a final time $T>0$. Without loss of generality, we can represent a pure state by $\rho(T) = \boldsymbol{e}_m \boldsymbol{e}_m^\dagger$, where $\boldsymbol{e}_m\in\R^N$ denotes the $m$-th unit vector in $\R^N$ such that the target $\rho(T)$ has one diagonal element being $1$ (at row and column $m$), and zeros everywhere else. 
For the specific application of the ground-state reset, the target density matrix can be formulated in this way with $m=0$, representing the ground state of the zero'th energy level.}

{
Note that considering the pure-state target be of the form $\boldsymbol{e}_m\boldsymbol{e}_m^\dagger$ is not a restriction, because a general pure state can be transformed into this form using a unitary basis transformation such that $U\rho U^\dagger = \boldsymbol{e}_m\boldsymbol{e}_m^\dagger$ for a unitary matrix $U$. In that case, the optimization can be performed in the basis defined by $U$, see Appendix \ref{app:general_pure_state}.
} 

The goal of the optimization is to drive any \textit{arbitrary} initial state $\rho(0)$ to the desired target $\boldsymbol{e}_m\boldsymbol{e}_m^\dagger$.
{We now }define a basis {for the vector space of all Hermitian matrices in $\C^{N\times N}$ spanning all possible initial states, that consists} of only density matrices. {We then introduce a specific} objective function $J$ that allows the basis {density} matrices to be lumped together in such a way that 
{the objective function needs to be evaluated for only one initial condition during the optimization process,}
hence reducing {the number of initial conditions that are to be considered}
from $N^2$ to $1$. 

{The $N^2$ density matrices, that span the vector space of all Hermitian matrices in $\C^{N\times N}$ over the field of real numbers, are defined as}
\begin{align} \label{eq:basismats}
B^{kj} := \frac 12 \left( \boldsymbol{e}_k \boldsymbol{e}_k^\dagger + \boldsymbol{e}_j \boldsymbol{e}_j^\dagger \right) +  \begin{cases} 
          0 & \text{if } \, k=j \\ 
        \frac 12 \left( \boldsymbol{e}_k \boldsymbol{e}_j^\dagger + \boldsymbol{e}_j \boldsymbol{e}_k^\dagger \right) & \text{if } \, k<j \\
        \frac i2 \left( \boldsymbol{e}_j \boldsymbol{e}_k^\dagger  - \boldsymbol{e}_k  \boldsymbol{e}_j^\dagger \right) & \text{if } \, k>j
      \end{cases}
\end{align}
for $k,j\in\{0,\dots, N-1\}$. {We note that each of these density matrices represents a pure state, $B^{kj} = \psib_{kj}\psib_{kj}^\dagger$.}
{In Appendix \ref{app:basismats_proof}, we prove that all} $B^{kj}$ are density matrices, {and that they are linearly independent in the vector space of Hermitian matrices in $\C^{N\times N}$ over $\R$, hence spanning all density matrices and providing a parameterization for density matrices (compare Appendix \ref{app:N2basismats}). Hence, any quantum state}  $\rho\in\C^{N\times N}$ can be written as a linear combination in this basis
\begin{align}\label{eq:arbitrarystate}
  \rho = \sum_{k,j=0}^{N-1} z_{kj} B^{k, j},
\end{align}
with coefficients $z_{kj}\in\R$. Naturally, since $\Tr(\rho)=1$, such coefficients satisfy $\sum_{kj}z_{kj}=1$. 

In contrast to other parameterizations of density matrices, such as the canonical basis for Hermitian matrices in $\C^{N\times N}$, or the Bloch-vector parameterization (compare Ref.~\onlinecite{bruning2012parametrizations}), the basis matrices $B^{kj}$ in \eqref{eq:basismats} are themselves density matrices representing quantum states. This is important because it ensures a physically meaningful 
{time evolution of the underlying quantum dynamics.}
As we shall see below, it also allows us to consider an {ensemble} of basis states as the only initial condition during {the optimization process for pure-state preparation}.

{To achieve that, we use the following objective function throughout the optimization towards the pure state $\boldsymbol{e}_m\boldsymbol{e}_m^\dagger$:}
{
\begin{align}\label{eq:objective_Jexpected}
       J(\rho(T)) &:= \Tr\left(N_m\rho(T)\right), 
\end{align}
where $N_m\in \R^{N\times N}$ is a diagonal matrix with diagonal elements $\lambda_i = |i-m|$ for all $i=0,\dots N-1$.}
{For the case of ground-state optimization ($m=0$), the objective function measures the expected energy level of the state at final time $T$, using the observable $N_0$. 
For pure-state preparation with $m>0$, the objective function measures a weighted sum of the population of all states except the $m$-th one. In both cases, minimizing $J$ drives the system to the state of the $m$-th energy level.}
 {Note that $\Tr\left( N_m \rho(T)\right)\geq 0$, with equality if and only if $\rho(T)=\boldsymbol{e}_m\boldsymbol{e}_m^\dagger$. Hence, $J=0$ if and only if $\rho(T)$ represents the desired pure target state $\boldsymbol{e}_m\boldsymbol{e}_m^\dagger$.}

{Using} the above objective function throughout the optimization process allows us to reduce the number of initial conditions that have to be considered to only one. We achieve this by defining the initial condition to be {the ensamble of the $N^2$ pure states $\psib_{kj}$, with equal probability}
as follows:
\begin{align}\label{eq:basis_superposition}
   \rho_s(0) := \frac{1}{N^2}\sum_{k,j=0}^{N-1} B^{kj}.
\end{align}
Evaluating the objective function \eqref{eq:objective_Jexpected} on the propagated state of this initial condition yields
\begin{align}
    J(\rho_s(T)) = J\left(\frac{1}{N^2}\sum_{kj}B^{kj}(T)\right) = \frac{1}{N^2} \sum_{kj} J\left(B^{kj}(T)\right),
\end{align}
due to the linearity of the solution operator of Lindblad's master equation with respect to the initial condition, as well as the linearity of $J$ with respect to $\rho(T)$. {Here,} $B^{kj}(T)$ denotes the propagated state corresponding to solving Lindblads master equation with initial condition $B^{kj}$.
Therefore, by minimizing $J(\rho_s(T))$ with one initial condition $\rho_s(0)$, we equivalently minimize the average {of $J$} over all basis density matrices. Further, if the optimum is achieved with $J(\rho_s(T))=0$, we get
\begin{align}
    {0 = }& {J(\rho_s(T)) =} 
     {\frac{1}{N^2} \sum_{k,j=0}^{N-1} \underbrace{\Tr\left(N_m B^{kj}(T)\right)}_{\geq 0}} \\
     \Rightarrow  \quad & J(B^{kj}(T))=0 \quad \forall \, k,j=0,\dots, N-1,
\end{align}
{such that the target is reached for each basis state. Since} any arbitrary initial quantum state at $t=0$ can be represented in this basis with $\rho(0) = \sum_{kj} z_{kj}B^{kj}$,
its propagated state at time $T$ {then} satisfies
\begin{align}
    J(\rho(T)) = \sum_{kj}z_{kj} J\left(B^{kj}(T)\right) = 0.
\end{align}
Hence, the system is in {the desired pure target state}. 
If {$J=0$} is not achieved during the optimization for the {ensemble} state $\rho_s$, a similar derivation yields the average error.

\section{Modeling a coupled qudit-cavity system}\label{sec:governingeq}

{We consider a quantum system consisting of a number of coupled qudits that also are coupled to a cavity resonator, where the qudits and the cavity interact with external control fields.}
{The }composite system {under consideration} consists of $Q-1$ qudits modeled with $n_q$ energy levels for the $q$-th subsystem ($q=1,\dots,Q-1$), coupled to a readout cavity modelled with $n_Q$ energy levels. 
{We make the standard assumptions that the quantum system interacts weakly with its environment (the bath), that there is no initial correlation between the system and the bath, and that the interaction between the system and the bath is Markovian. These assumptions lead to Lindblad’s master equation~\cite{breuer2002theory} governing the time-evolution of the density matrix describing the quantum system, $\rho \in \C^{N\times N}$, with dimension $N := \prod_{q=1}^Q n_q$: }
\begin{align}\label{eq:mastereq_lab}
  \frac{\mathrm{d}\,\rho(t)}{\mathrm{d} \,t} = &-i\left[H(t), \rho(t)\right] + \Ell\left(\rho(t)\right), \quad t \in (0,T).
\end{align}
{Here,} $H(t)$ denotes the Hamiltonian describing the system and its controls, the commutator operator is defined by $[A,B]=AB - BA$, and $\Ell\left(\rho\right)$ denotes the Lindbladian operator that models interactions between the quantum system and its environment, as specified below.

The Hamiltonian is decomposed into a time-independent system part ($H_d$) and a time-varying control part ($H_c(t)$) that models the action of external control fields: $H(t) = H_d + H_c(t)$. While not a restriction of our approach, in this paper we exemplify the techniques on a typical circuit QED system within the dispersive coupling regime with Hamiltonians of the form (see e.g. Ref.~\onlinecite{PhysRevLett.108.240502})
\begin{align}\label{eq:drifthamiltonian}
  H_d = \sum_{q=1}^Q \left(\omega_q a_q^\dag a_q - \frac{\xi_q}{2} a_q^{\dagger}a_q^{\dagger}a_q a_q - \sum_{p>q} \xi_{pq} a_p^{\dagger}a_p a_q^{\dagger} a_q  \right). 
\end{align}
Here, $\omega_q$ and $\xi_q$ denote the ground state transition frequency and self-Kerr coefficient of sub-system $q$; the cross-Kerr coefficient between subsystems $p$ and $q$ is denoted $\xi_{pq}$. Furthermore, 
$a_q$ denotes the lowering operator for subsystem $q$,
\begin{align}
  a_q := I_{n_1} \otimes \dots \otimes I_{n_{q-1}} \otimes A_{n_q} \otimes I_{n_{q+1}}\otimes \dots \otimes I_{n_Q} \, \in \R^{N\times N},
  \end{align}
where $I_{n_q}$ denotes the identity matrix in $\R^{n_q \times n_q}$ and the one-dimensional lowering operator satisfies
\begin{align}
    A_{n_q} := \begin{pmatrix}
    0 & \sqrt{1}     \\
     &  \ddots   & \ddots  \\
     &           &  \ddots & \sqrt{n_q-1}  \\
     &           &         & 0   
 \end{pmatrix} \in \R^{n_q \times n_q}.
\end{align}

{In the computational basis, we represent the ground state, $|0\rangle$, by the first unit vector in $\R^N$, denoted by $\boldsymbol{e}_0$. Similarly, the $j^{th}$ excited state $|j\rangle$ is represented by the unit vector $\boldsymbol{e}_j\in\R^N$}. 

The action of external control fields on the quantum system is modelled through the control Hamiltonian,
\begin{align}\label{eq_hamctrl}
   H_c(t) := \sum_{q=1}^Q f^q(\vec{\alpha}^q,t) (a_q + a_q^{\dagger}),
\end{align}
with real-valued, time-dependent control functions $f^q(\vec{\alpha}^q,t)$ that are parameterized by real-valued parameters $\vec{\alpha}^q\in \R^d$, which are to be determined through optimization. 

The Lindbladian operator $\Ell\left(\rho(t)\right)$ is assumed to be of the form
\begin{align}
    \Ell\left(\rho\right) =  \sum_{q=1}^Q \sum_{l=1}^2  \Ell_{lq} \rho \Ell_{lq}^{\dagger} - \frac 1 2 \left( \Ell_{lq}^{\dagger}\Ell_{lq} \rho + \rho\Ell_{lq}^{\dagger} \Ell_{lq}\right).
\end{align}
Here, the collapse operators $\Ell_{lq}$ model decay and dephasing processes in subsystem $q$ with $\Ell_{1q} := \frac{1}{\sqrt{T_1^q}} a_q$ (decay) and $\Ell_{2q} :=  \frac{1}{\sqrt{T_2^q}} a_q^{\dagger}a_q$ (dephasing). 
The positive constants $T_1^q$ and  $T_2^q$ correspond to the decay and dephasing times on subsystem $q=1,\dots,Q$, respectively.

{
In order to slow down the time-scale of Lindblad's master equation, we employ the rotating wave approximation with frequencies $\omega_q$, cancelling out the first term in the system Hamiltonian \eqref{eq:drifthamiltonian}. 
}

{In summary, the optimization problem for pure-state preparation becomes}
\begin{align}
\min \; &J(\rho(T)) \nonumber\\
 \text{s.t.} \quad \frac{\mathrm{d}\rho}{\mathrm{d} t}&= -i \left[H(t), \rho\right] + \Ell(\rho), \quad \forall\; t\in (0,T) \label{eq:optim_oneinitial}\\
 \rho(0) &= \rho_s(0).\nonumber
\end{align}
{Note, that only the initial condition $\rho_s(0)$ needs to be propagated by Lindblad's master equation.}

\section{Numerical Approach}\label{sec:numericalapproach}

{The general framework as presented in Section \ref{sec:optim} (the proposed basis elements in \eqref{eq:basismats}, the objective function \eqref{eq:objective_Jexpected} and the ensemble initial state in \eqref{eq:basis_superposition}) could in principle be utilized in existing numerical optimization methods for quantum control, such as Grape or Krotov.}
{However, in this paper we present and employ an alternative approach to numerically solving the open system optimization problem \eqref{eq:optim_oneinitial}, as implemented in the open-source software package Quandary \cite{quandaryGithub} targeting high-performance computing architectures.
}

{We apply} iterative gradient-based updates to the control pulses {in order to solve the} optimization problem \eqref{eq:optim_oneinitial}, preconditioned by a Hessian approximation using L-BFGS updates \cite{nocedal2006numerical}. 
In each iteration of the optimization process, Lindblad's master equation is solved numerically {in the rotating frame} to propagate $\rho_s(0)$ to $\rho_s(T)$. {To do so, we employ} a second-order implicit time-integration scheme (Implicit Midpoint Rule, IMR\cite{hairer2006geometric}) on an equidistant time grid $t_i = i\Delta t$ for $i=0,\dots,N_T$ with step size $\Delta t>0$ and $T=N_T\Delta t$. 
The Implicit Midpoint Rule is a Runge-Kutta time-integration scheme that is symplectic, hence avoiding numerical (artificial) dissipation throughout the numerical solution process. 

The real-valued laboratory frame control function can be written as
\begin{align}\label{eq:control_lab}
    f^q(\vec\alpha^q, t) &= 2\,\mbox{Re}\left( d^q(\vec\alpha^q, t) e^{i\omega_q\, t}\right).
\end{align} 
for computational rotating-frame control functions $d^q(\vec{\alpha}^q, t)$. 
{W}e parameterize the {rotating-frame} controls 
using $N_s$ fixed B-spline basis functions that act as the envelope for $N_f$ carrier waves:
\begin{align}\label{eq_rotctrl}
  d^q(\vec{\alpha}^q,t) &= \sum_{s=1}^{N_s}S_s(t) \sum_{n=1}^{N_f}  \alpha^{q}_{s,n}\,e^{it\Omega_q^n},
\end{align}
where $\alpha^{q}_{s,n} = \alpha^{q (1)}_{s,n} + i \alpha^{q (2)}_{s,n} \in \C$ are the control amplitudes that are to be determined through optimization, giving a total of $2N_s N_f$ real-valued optimization parameters per subsystem $q$. 
The basis functions $S_s(t)$ are chosen to be piece-wise quadratic B-spline wavelets\cite{Unser97}, which have local support in time {and are continuously differentiable}. The wavelets are centered at times $\tau_s = \Delta \tau (s - 3/2)$ for $s=1,2,\ldots,N_s$ where the knot spacing is $\Delta \tau = T/(N_s-2)$. 
Further, $\Omega_q^n\in\R$ denote the carrier wave frequencies in the rotating frame.

{B-splines with carrier waves provide a compact alternative to discretizing the control functions on the same time step as Lindblad's master equation, because the frequencies of the carrier waves can be chosen to be focused near the resonance frequencies of the quantum system, triggering transitions between the energy levels\cite{petersson2021optimal}.}
By substituting \eqref{eq_rotctrl} into \eqref{eq:control_lab}, the lab-frame carrier frequencies become $\omega_q + \Omega_q^n$. 
Those frequencies {are} chosen to match the transition frequencies in the system Hamiltonian \eqref{eq:drifthamiltonian}. For example, when $\xi_{pq}\ll \xi_q$, the lab frame transition frequency between energy levels $n-1$ and $n$ in subsystem $q$ satisfies $\omega_q - (n-1)\xi_q$. Thus, by choosing $\Omega_q^n = - (n-1)\xi_q$ we trigger transition between energy levels $n-1$ and $n$ in subsystem $q$. 
{The B-splines modulate the amplitude and phase of each carrier wave. Because they vary on a much slower time scale than the carrier waves, the number of B-spline parameters can be significantly smaller than the number of time steps.
In contrast to other control parameterizations using basis functions (such as Fourier modes or Legendre polynomials, compare e.g. Ref.~\onlinecite{caneva2011chopped}), each B-spline wavelet is local in time. As a result, each B-spline coefficient only influences the envelope function locally. In numerical experiments we have found that this ''time-local`` support of the control parameterization results in a more regular optimization surface with fewer local minima than when the same problem is evaluated using a fully delocalized parameterization (e.g. a Fourier basis).} 

We follow the \textit{first-discretize-then-optimize} approach, and compute the gradient {of the objective function with respect to the B-spline coefficients} using the discrete adjoint approach. Since the IMR is a symmetric Runge-Kutta scheme, the discrete adjoint time-stepping scheme to solve the adjoint equation is again the IMR, now propagating sensitivities backwards through the time domain while collecting contributions to the gradient at each time-step. 
The discrete adjoint approach yields an exact gradient on the discrete level, at a computational cost that is independent of the number of control parameters. 

In order to stabilize the optimization, we employ a Tikhonov regularization, adding the convex term 
\begin{align}
    \gamma_1 \|\bfa\|^2_2
\end{align}
to the objective function, for a parameter $\gamma_1 >0$ and {the control vector} $\bfa = \left( \vec{\alpha}^1,\dots, \vec{\alpha}^Q \right)$.
Additionally, an integral term can be added that penalizes {the final-time objective} over time with 
\begin{align}
    \gamma_2\int_0^T w(t) J\left(\rho(t)\right) \, \mathrm{d}t,
\end{align}
where $w(t) = \frac{1}{a}\exp\left(-\left(\frac{t-T}{a}\right)^2\right)$ is a weight function and $a>0$ is a tunable parameter. This penalty term drives $\rho(t)$ towards the target state near the final time.

\section{Numerical results}\label{sec:numerical_results}

{This section presents numerical results utilizing the framework in Section \ref{sec:optim} and the numerical tools in Section \ref{sec:numericalapproach}. We demonstrate the shaping of control pulses for 
unconditional ground-state reset of one and two qubits, coupled to a readout cavity. As noted in Section \ref{sec:optim} as well as in Appendix \ref{app:general_pure_state}, demonstrating pure-state preparation for the ground state $\boldsymbol{e}_0\boldsymbol{e}_0^\dagger$ generalizes naturally to any other pure target states. A numerical example of the latter is presented in Appendix~\ref{app:results_purestateprep}.}

{First, we consider one qudit} modelled with $n_1 = 3$ energy levels {that is coupled to a cavity modelled with $n_2 = 20$ energy levels, such that $\rho\in\C^{N\times N}$ with $N=60$.}
Table \ref{tab:3x20_coeffs} lists the system parameters, such as transition frequencies $\omega_q$, anharmonicity $\xi_q$, and decoherence times of the qudit, and the cavity, as well as the dispersive cross-Kerr coupling $\xi_{12}$. 
{The chosen system parameters are drawn from a specific experimental platform currently under study at Lawrence Livermore National Laboratory and are intended to be representative of typical present-day superconducting circuits.}
Note that decoherence is significantly faster in the cavity than in the qudit.

\begin{table}
    \caption{\label{tab:3x20_coeffs} System parameters for reset of a qudit in a cavity.}
    \begin{ruledtabular}
        \begin{tabular}{llcccc}
              &&  $\omega_q/2\pi$ [GHz] & $\xi_q / 2\pi$ [MHz] & $T_{1,q}$ [$\mu$s] & $T_{2,q}$[$\mu$s] \\
            \hline
            & Qudit ($q=1$)  & $4.41666$  & $230.56$   & $80$     & $26$  \\
            & Cavity ($q=2$) & $6.84081$  & $0$      & $0.3892$ & --\\
            \hline
            & cross-Kerr&        & $1.176$ && \\
        \end{tabular}
    \end{ruledtabular}
\end{table}

{We first} assume the cavity to be in its ground state at time $t=0$ {(this assumption will later be dropped).} 
The basis for the density matrix {therefore only needs to span all qudit states. In this case, the initial condition that is used throughout the optimization process becomes}
\begin{align}\label{eq:3x20_initcond}
    \rho_s(0) = \frac{1}{3^2}\sum_{k,j=0}^2 B^{kj} \otimes {\bf e}_0{\bf e}_0^\dag 
\end{align}
{with $B^{kj}\in \C^{3\times 3}$ defined in \eqref{eq:basis_superposition} in the qudit's space dimensions, and where} ${\bf e}_0\in \R^{20}$ denotes the first unit vector in the cavity's dimension.

{The optimization target aims to drive any initial qudit state to the ground state in $T=2.5\mu$s, while leaving the cavity empty at the final time.} The objective function {reads 
\begin{gather}
    J(\rho(T)) = \Tr\left(N_0\rho(T)\right), \\
    \text{with} \quad N_0 = \begin{bmatrix} 0  \\  & 1   \\ & & 2 \\ & & & \ddots \end{bmatrix} \in \R^{60\times 60},
\end{gather}}
where $\rho(T)$ solves Lindblad's master equation for the above initial condition. {Note that $J$ measures the expected energy level for the coupled qudit-cavity system.}

\begin{figure}
    \centering
    \input{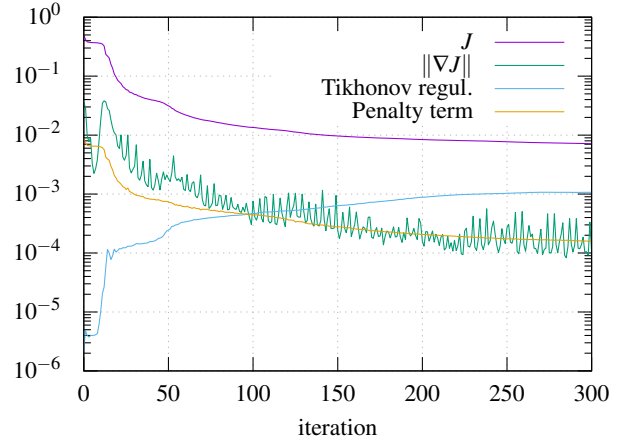}
    \caption{Optimization history for optimal reset of a qudit in a cavity.}
    \label{fig:3x20_sumAlice_optimhistory}
\end{figure}

Each control function is parameterized by $N_s=75$ spline basis functions. The carrier wave frequencies in the rotating frame are chosen as $\Omega_{q=1}^1 = 0$ and $\Omega_{q=1}^2 = -\xi_1$ for the qudit and $\Omega_{q=2}^1 = 0$ for the cavity.  
In order to conform with control hardware limitations, bounds for the lab frame control amplitudes on the qubit are included in the optimization process by applying a projected line-search, with maximum amplitude of $36/2\pi$ {MHz}.
We choose a time-step size of $\Delta t=1\cdot 10^{-4}\,\mu$s corresponding to $N_t=25,000$ time steps, to numerically integrate Lindblad's master equation in the rotating frame. 
We add the Tikhonov regularization term to the objective function with $\gamma_1=10^{-6}$ as well as the weighted integral penalty term for penalizing the expected energy level over time with $\gamma_2=10^{-2}$ and $a=0.1$. Figure \ref{fig:3x20_sumAlice_optimhistory} shows the optimization progress of the L-BFGS scheme, demonstrating successful optimization in terms of a monotone decrease in the objective function and a relative drop in the gradient norm by two orders of magnitude. 

\begin{figure}
    \centering
    \input{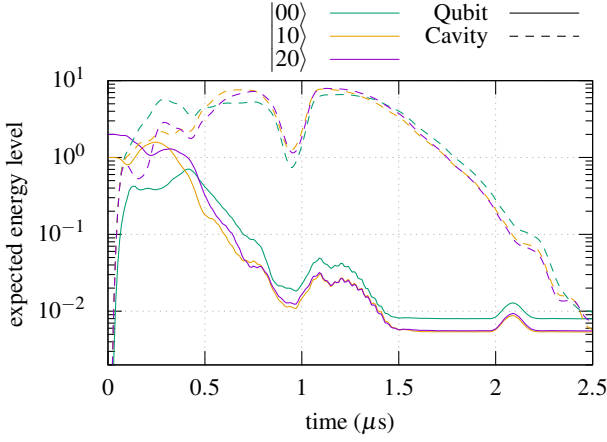}
    \caption{Optimal qudit reset: Evolution of expected energy level of the qubit (solid lines) and the cavity (dashed lines) for initial qubit states $|0\rangle,|1\rangle,|2\rangle$, the cavity starts in the ground state. Average fidelity at $T=2.5$us: $99.50\%$ (qubit), $99.37\%$ (cavity).}
    \label{fig:GS_expected_basis}
\end{figure}

To demonstrate the efficacy of the optimized control pulses on arbitrary initial qudit states, the optimized evolution of the expected energy level for the qudit and the cavity are shown in Figure \ref{fig:GS_expected_basis}
for the first three excited initial qudit states. 
Interestingly, the different initial states quickly collapse to about the same expected energy level, before evolving to the ground state at time $T=2.5\,\mu$s. 
{At that time, an average ground state fidelity\footnote{{The average fidelity is based on the fidelity as defined in Ref.~\onlinecite{Nielsen-Chuang}, 
\begin{align*}
    F(\sigma, \rho) := \left(
    \Tr \sqrt{\sqrt{\sigma}\rho\sqrt{\sigma}}\right)^2,
\end{align*}
between the pure state target $\sigma = \boldsymbol{e}_m\boldsymbol{e}_m^\dagger$ and the realized state $\rho$, averaged over all basis elements $\rho = B^{kj}(T)$ at the final time $T$: 
\begin{align*}
    F_{avg} := \frac{1}{N^2}\sum_{kj} F\left(\boldsymbol{e}_m\boldsymbol{e}_m^\dagger, B^{kj}(T) \right) = \frac{1}{N^2}\sum_{kj} \left[B^{kj}(T)\right]_{mm}
\end{align*}
}}
of $99.50\%$ and $99.37\%$ is reached for the qudit and the cavity, respectively.} 

\begin{figure}
    \centering
    \input{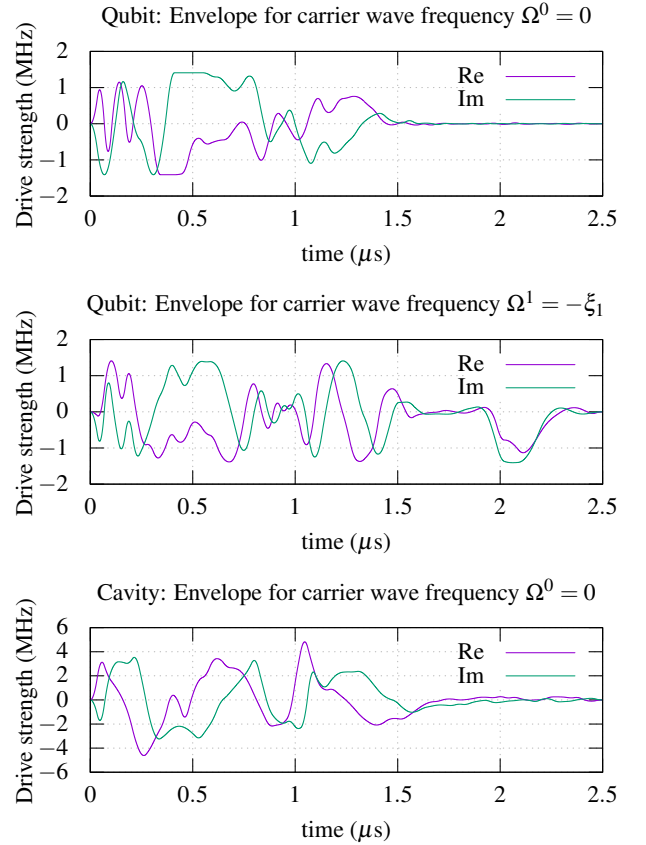}
    \caption{Optimized rotating frame control pulses: B-spline envelopes for each carrier wave frequency driving the qubit and the cavity.}
    \label{fig:3x20_sumAlice_optcontrol}
\end{figure}

\begin{figure}
    \centering
    \includegraphics[width=.22\textwidth]{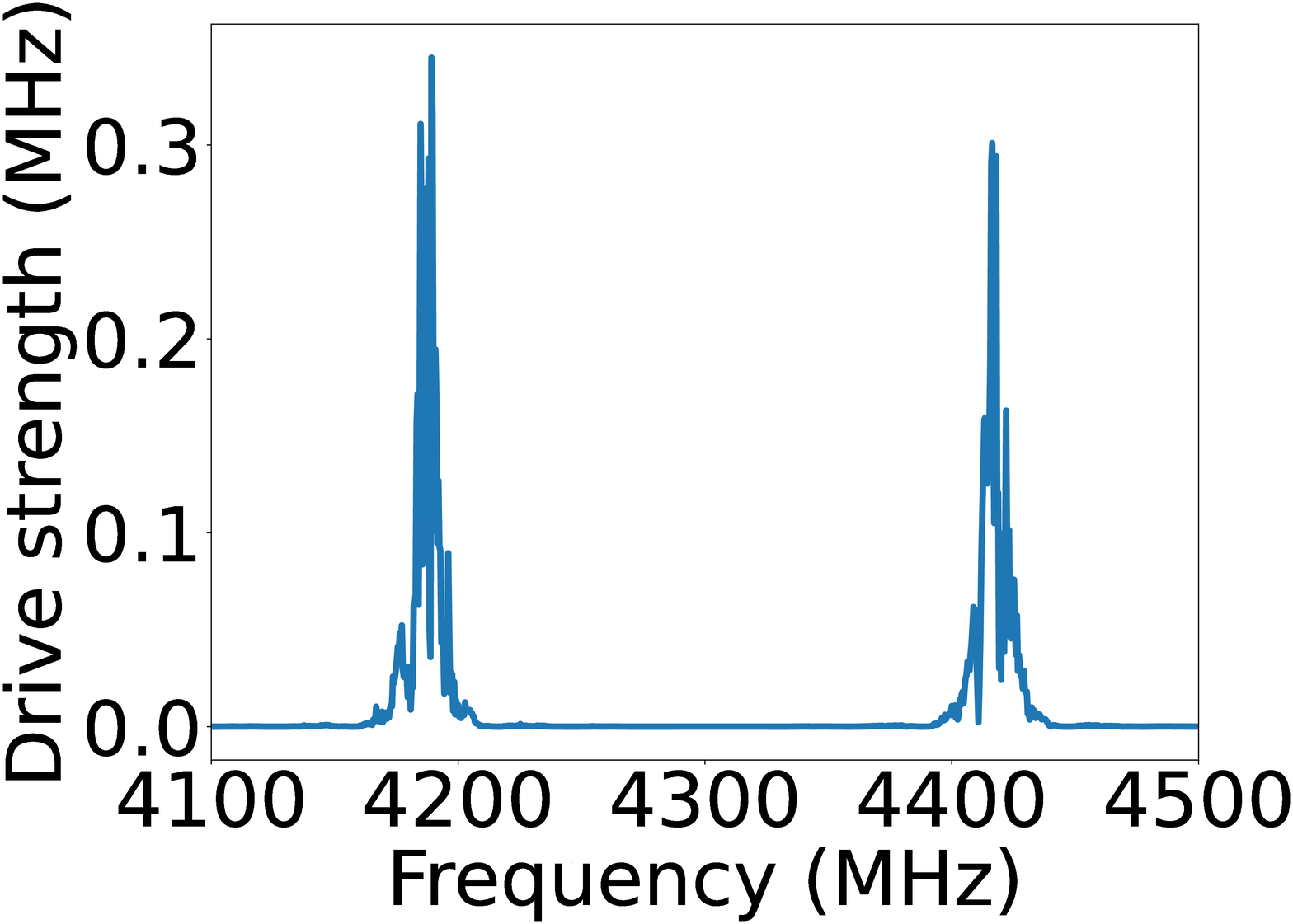}
    \includegraphics[width=.22\textwidth]{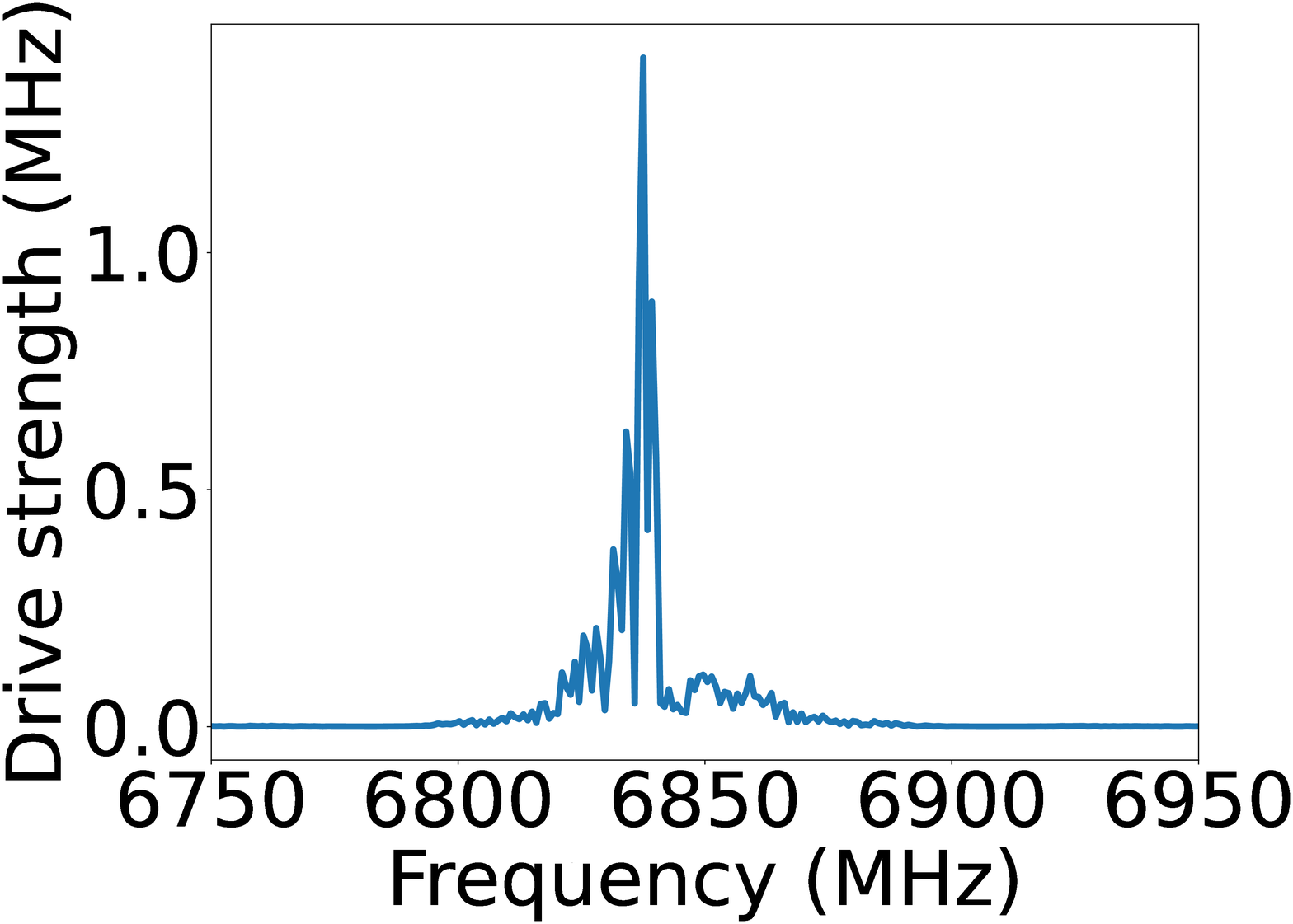}
    \caption{Fourier spectrum of the optimized control pulses for the qudit (left) and the cavity (right).}
    \label{fig:3x20_sumAlice_optcontrol_freq}
\end{figure}

The optimized rotating frame control pulses are visualized in Figure \ref{fig:3x20_sumAlice_optcontrol}, in terms of the B-spline envelopes for each carrier wave frequency.
The Fourier spectrum of the resulting lab-frame control pulses driving the qudit and the cavity, $f^1(\vec{\alpha}^1_{opt},t) $ and $f^2(\vec{\alpha}^2_{opt},t)$, are shown in Figure \ref{fig:3x20_sumAlice_optcontrol_freq}. It clearly visualizes how the carrier wave frequencies precisely trigger the underlying system frequencies. 

\begin{figure}
    \centering
    \input{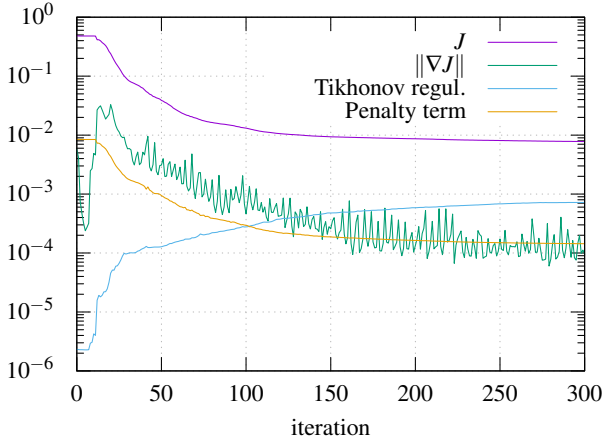}
    \caption{Optimization history for simultaneous reset of a qubit and a cavity.}
    \label{fig:fullbasis_optimhistory}
\end{figure}

{In general, the cavity may have thermal occupation at time $t=0$. In our next numerical example, we therefore
consider any initial coupled qudit-cavity state.} 
The initial conditions considered for optimization is {hence} the {ensemble} of basis density matrices in the full dimensions $B^{kj} \in C^{60\times 60}$:
\begin{align}
    \rho_s(0) = \frac{1}{60^2} \sum_{k,j=0}^{59} B^{kj},
\end{align}
{for $B^{kj}\in \C^{60\times 60}$ as in \eqref{eq:basis_superposition}.}
To account for the increased complexity of the optimization landscape, we increase the duration of the control pulses to $T=4\mu s$, and allow for $N_s=100$ spline basis function to parameterize the controls. 
The optimization history, shown in Figure \ref{fig:fullbasis_optimhistory}, demonstrates similar convergence behavior as in the previous test case. 

\begin{figure}
    \centering
    \input{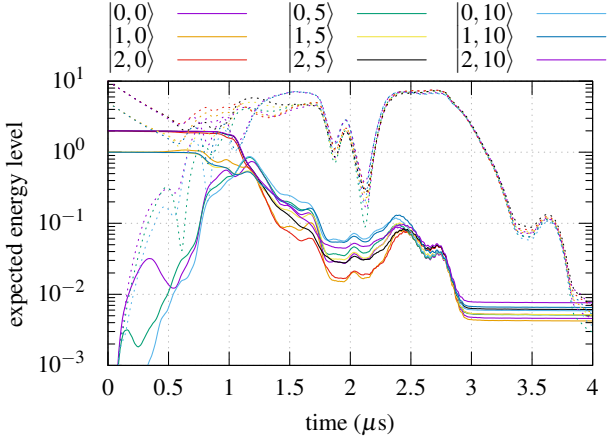}
    \caption{Optimized evolution of expected energy levels of the qudit (solid lines) and the cavity (dashed lines) for various initial states. Average fidelity at $T=4\mu s$: $99.30\%$ (qudit), $99.29\%$ (cavity).}
    \label{fig:fullbasis_expected}
\end{figure}

\begin{figure}
    \centering
    \input{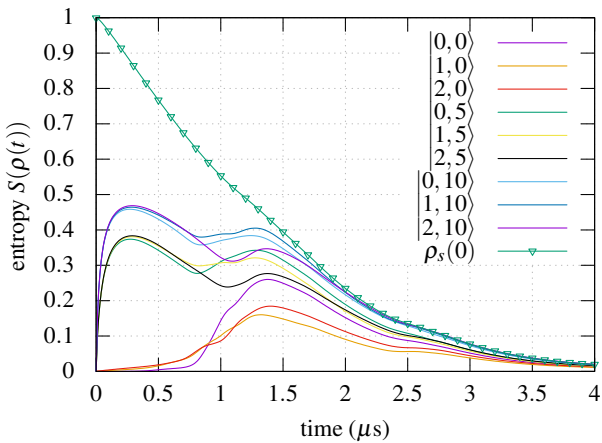}
    \caption{{Evolution of the von Neumann entropy for various initial states, driven by the optimized control pulses.}}
    \label{fig:entropy}
\end{figure}

Utilizing the optimized controls, the evolution of the expected energy levels for various initial qudit and cavity states are shown in Figure \ref{fig:fullbasis_expected}. At $T=4us$, an average fidelity of $99.37\%$ ($99.15\%$) for the qudit (cavity) is attained.
{To gain further insight into the purification process, we evaluate the evolution of the (normalized) von Neumann entropy, 
\begin{align}
    S(\rho(t)) = -\frac{1}{\log(N)}\Tr\left(\rho(t) \log \left( \rho(t) \right)\right).
\end{align}
Figure \ref{fig:entropy} shows monotonic decrease of the entropy of the maximally mixed initial ensemble state $\rho_s(0)$, towards the pure ground-state target. We also show the evolution of the entropy starting from various initial pure states. In that case, the entropy is initially zero, ramps up intermittently, before it finally decays towards zero. However, it always remains bounded by the entropy of the maximally mixed initial state. This indicates that the maximally mixed initial state represents a "worst case" scenario. This is what the optimizer uses to shape the control functions, rather than any particular initial state. Thus we may think of the optimized control pulses as the most efficient way of reducing the entropy in the system, averaged over all initial states.
}

{We note that by utilizing the sole initial condition $\rho(0) = \rho_s(0)$,
the computational {burden} for each evaluation of the objective function is reduced by a factor of $60^2=3600$.}

\begin{figure}
    \centering
    \includegraphics[width=.49\textwidth]{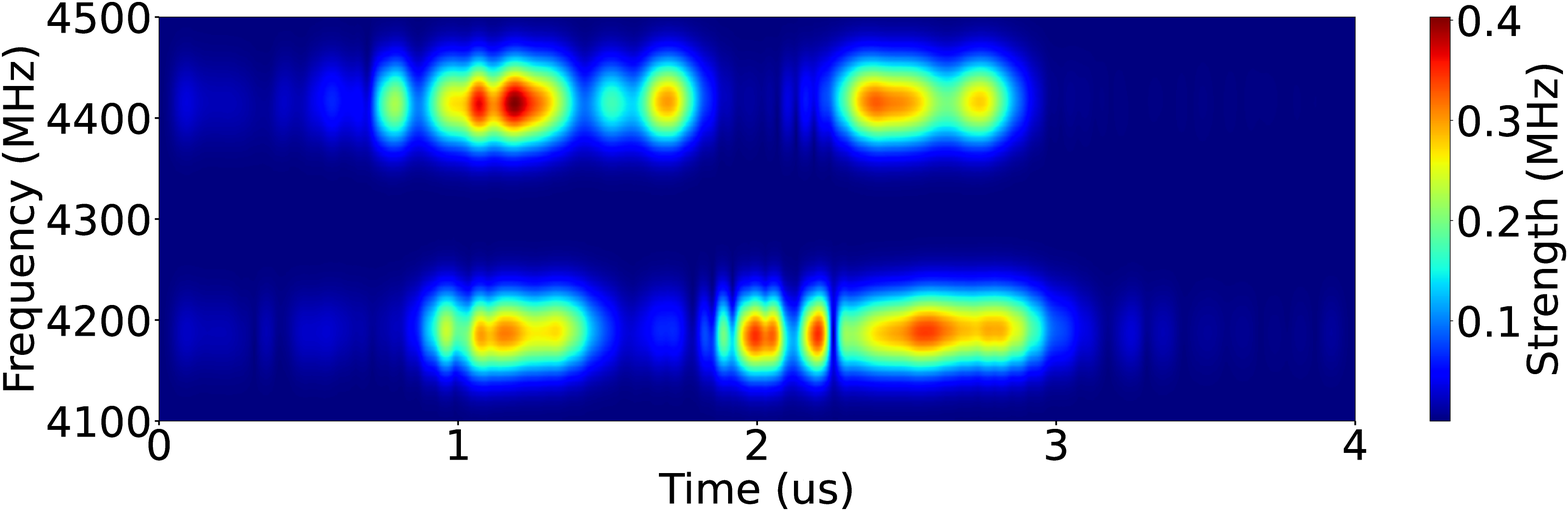}
    \includegraphics[width=.49\textwidth]{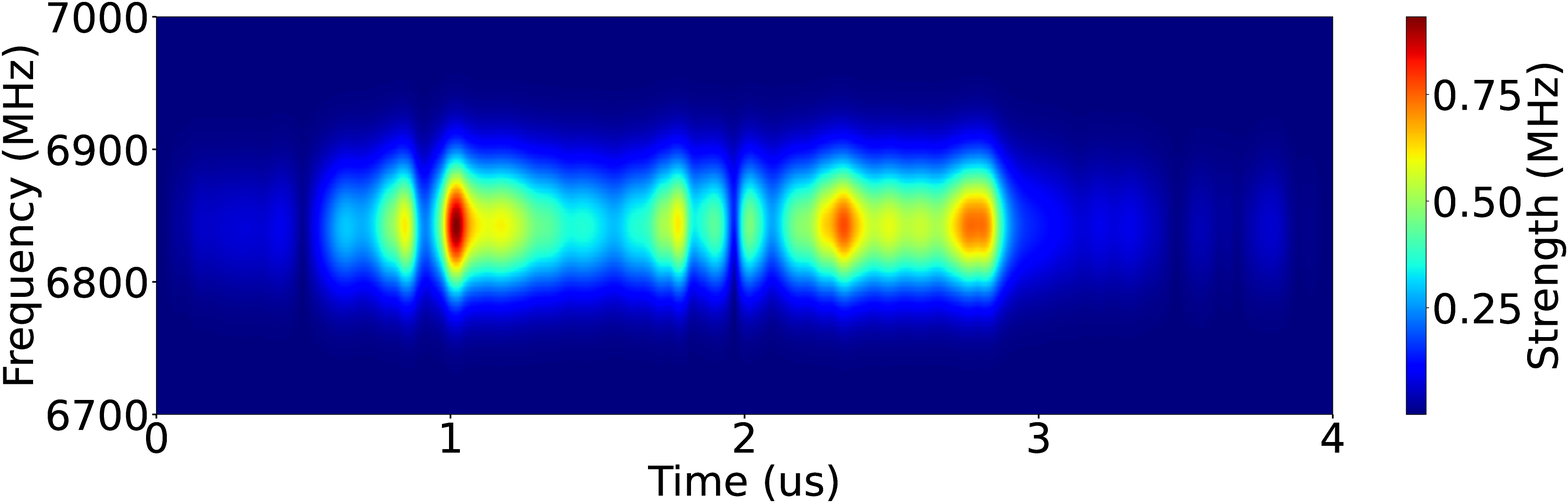}
    \caption{Spectogram of optimized controls for the qudit (top) and cavity (bottom).}
    \label{fig:fullbasis_spectogram}
\end{figure}

The optimal control pulses driving the qudit and cavity are visualized in Figure \ref{fig:fullbasis_spectogram} in terms of their spectogram\footnote{{Spectograms are computed by a Morlet wavelet analysis\cite{GOUPILLAUD198485} with $150$ cycles of the control signal, using a sample rate of $1/\Delta t$ (here, $\Delta t$ is the time-step size in the time-integration scheme)}}, where the color indicates drive strength.
Again, it is visible how the carrier waves in the control pulse parameterization trigger the resonant frequencies in the system. In addition, low control amplitudes during the first and last microsecond indicate that further optimization with a shorter duration might be achievable. 

{In order to demonstrate the applicability of the proposed framework to more complex systems, we now consider ground-state reset of two qubits coupled to a readout cavity. We model the two qubits with $n_1=n_2=2$ and the cavity with $n_3=20$ energy levels}. We consider a fully coupled system with parameters given in Table \ref{tab:2x2x20_coeffs}. We aim to drive the coupled system from any initial state  
to the ground state in $T=5\mu s$, discretized with a time step of $\Delta t=10^{-4}\mu s$. 
The initial condition spans the basis of the coupled qubit-qubit-cavity system with 
\begin{align}
    \rho_s(0) =  \frac{1}{80^2} \sum_{k,j=0}^{79} B^{kj},
\end{align}
where the basis density matrices $B^{kj}\in \C^{80\times 80}$.

\begin{table}
    \caption{\label{tab:2x2x20_coeffs} System parameters for reset of two qubits in a cavity.}
    \begin{ruledtabular}
        \begin{tabular}{llcccc}
              &&  $\omega_q/2\pi$ [GHz] & $\xi_q / 2\pi$ [MHz] & $T_{1,q}$ [$\mu$s] & $T_{2,q}$[$\mu$s] \\
            \hline
            & Qubit 1  & $4.41666$  & $230.56$   & $80$     & $26$  \\
            & Qubit 2  & 4.510  & 251.0   & $90$     & $30$  \\
            & Cavity & $6.84081$  & $0$      & $0.3892$ & $0$\\
            \hline
            & Q1$\leftrightarrow$ Q2 &     & $0.001$ && \\
            & Q1$\leftrightarrow$ Cav.&    & $1.176$ && \\
            & Q2$\leftrightarrow$ Cav.&    & $1.2$ && \\
        \end{tabular}
    \end{ruledtabular}
\end{table}

\begin{figure}
    \centering
    \input{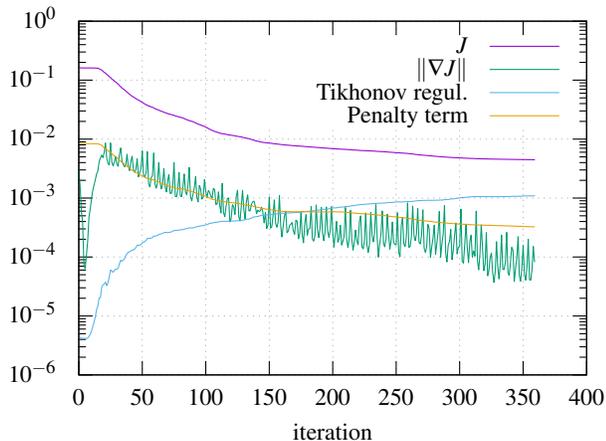}
    \caption{Simultaneous reset of two qubits: Optimization history.}
    \label{fig:2x2x20_optimhistory}
\end{figure}

The control functions are parameterized with $100$ B-spline wavelets, with carrier waves $\Omega^1_{q=1} =  \Omega^1_{q=2} = \Omega_{q=3}^1 = 0$, giving a total of $1,000$ real-valued optimization variables. Tikhonov regularization and penalty parameters are added as in the previous test cases.
The optimization history is plotted in Figure \ref{fig:2x2x20_optimhistory}, demonstrating the robustness of optimization progress with respect to the increased complexity of this test problem. 

\begin{figure}
    \centering
    \input{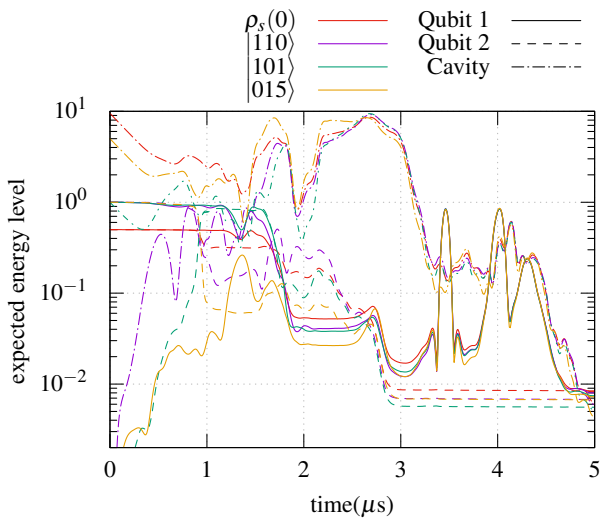}
    \caption{Optimal reset of two qubits and cavity: Optimized evolution of expected energy levels for the initial {ensemble} state $\rho_s(0)$ and various other initial states. At $T=5\mu s$, the system is in its ground state with average fidelity of $99.27\%$ (qubit 1), $99.47\%$ (qubit 2), and $99.55\%$ (cavity).}
    \label{fig:2x2x20_expected}
\end{figure}

Applying the optimized controls to various initial conditions, Figure \ref{fig:2x2x20_expected} shows the evolution of the expected energy levels for each subsystem over time. At final time $T=5\mu s$, the average ground-state fidelity is $99.18\%$ for the first qubit, $99.16\%$ for the second qubit, and $99.4\%$ for the cavity. 

\section{Conclusion}

This paper presents a numerical optimal control framework for shaping control pulses to drive a composite {open} quantum system {to a pure target state.}
Typically a basis of initial states needs to be considered {throughout the optimization} in order to {account for} all possible initial quantum states. We develop a basis for the vector space of Hermitian matrices in such a way that each basis {element} is a density matrix itself. 
{Using an ensemble of those basis states together with a specific objective function allows the number of initial conditions to be reduced to only one.}
{This reduction is particularly significant when considering that each solve of Lindblad's master equation itself scales as $O(N^2)$.}

We solve the resulting optimal control problem by applying an adjoint-based, preconditioned gradient descent algorithms (L-BFGS), solving Lindblad's master equation {for the ensemble state} in each optimization iteration. The control pulses are parameterized using a set of smooth B-spline basis functions that act as envelopes to carrier waves, precisely triggering resonant frequencies in the system.
We demonstrate the effectiveness {o}f the approach {by numerical examples for unconditional ground-state} reset of one and two qubits in a cavity, {as well as the preparation of the $|10\rangle$ state}. 

Future work will include experimental validation of the optimized control pulses on a real-world quantum device.

\begin{acknowledgments}
This work was performed under the auspices of the U.S.\ Department of Energy by Lawrence Livermore National Laboratory under Contract DE-AC52-07NA27344 (LLNL-JRNL-823510).\\
This document was prepared as an account of work sponsored by an agency of the United States government.
Neither the United States government nor Lawrence Livermore National Security, LLC, nor any of their employees makes any warranty, expressed or implied, or assumes any legal liability or responsibility for the accuracy, completeness, or usefulness of any information, apparatus, product, or process disclosed, or represents that its use would not infringe privately owned rights.
Reference herein to any specific commercial product, process, or service by trade name, trademark, manufacturer, or otherwise does not necessarily constitute or imply its endorsement, recommendation, or favoring by the United States government or Lawrence Livermore National Security, LLC.
The views and opinions of authors expressed herein do not necessarily state or reflect those of the United States government or Lawrence Livermore National Security, LLC, and shall not be used for advertising or product endorsement purposes.
\end{acknowledgments}

\section*{Data Availability Statement}

The data that support the findings of this study are openly available in the software package Quandary\cite{quandaryGithub}.

\appendix

\section{Proof that $B^{kj}$ are linearly independent density matrices in $\C^{N\times N}$.}\label{app:basismats_proof}

{Here, we prove that the $N^2$ matrices $B^{kj}\in \C^{N\times N}, k,j=0,\dots, N-1$ defined in \eqref{eq:basismats} are density matrices, i.e. they are Hermitian, positive semi-definite, with unit trace, and that they are linearly independent in the vector space of Hermitian matrices in $\C^{N\times N}$ over $\R$.}

  It is easy to see that all $B^{kj}$ are Hermitian, and that $\Tr(B^{kj}) = 1$ for all $k,j=0,\dots,N-1$. First we show that all $B^{kj}$ are positive semi-definite, proving that all $B^{kj}$ are density matrices. 
  
  Let $x\in \C^N$. For all $k=0,\dots,N-1$, we have
  \begin{align}
     x^\dag B^{kk} x &= \bar x_k x_k = |x_k|^2 \geq 0
  \end{align}
  Further, for all $0\leq k<j\leq N-1$:
  \begin{align}
         x^\dag B^{kj} x &= \frac{1}{2} \left(x^\dag E^{kk}x + x^\dag E^{jj}x + x^\dag E^{kj}x + x^\dag E^{jk}x \right) \nonumber \\
         &=\frac{1}{2} \left( |x_k|^2 + |x_j|^2 + 2\mbox{Re}(\bar x_k x_j) \right) \nonumber \\
         &= \frac 12 |x_k + x_j |^2 \geq 0,
  \end{align}
  where $E^{kj}:= \boldsymbol{e}_k\boldsymbol{e}_j^\dagger$.
  Lastly, for all $0\leq j<k\leq N-1$: 
  \begin{align}
     x^\dag B^{kj} x &= \frac{1}{2} \left(x^\dag E^{kk}x + x^\dag E^{jj}x + ix^\dag E^{kj}x - ix^\dag E^{jk}x \right) \nonumber \\
         &=\frac{1}{2} \left( |x_k|^2 + |x_j|^2 + 2i\, \mbox{Im}(\bar x_k x_j) \right) \nonumber \\
         &=\frac{1}{2} | x_k + i x_j |^2 \geq 0.
  \end{align}

   Next, we show that all basis elements are linearly independent in the vector space of all Hermitian matrices.
   To this end, let 
  \begin{align}\label{eq:linearindependent}
      \textbf{0} = \sum_{k=0}^{N-1} z_{kk} B^{kk} + \sum_{k=0, j=k+1}^{N-1,N-1} z_{kj} B^{kj} + \sum_{j=0, k=j+1}^{N-1,N-1} z_{kj} B^{kj}
  \end{align}
  for real-valued coefficients $z_{kj} \in \R$. We proceed by proving that $z_{kj} = 0$ for all $k,j$. First, consider an off-diagonal element in the null-matrix $\textbf{0}$ at position $(m,n)$ with $m\neq n$. For this element, Equation~\eqref{eq:linearindependent} reads
  \begin{align}
      0 &=  \frac 12 z_{mn} + \frac i2 z_{nm}, \quad \text{if} \quad m>n\\
      0 &=  \frac 12 z_{mn} - \frac i2 z_{nm}, \quad \text{if} \quad m<n.
  \end{align}
  It follows that $z_{mn} = z_{nm} = 0$. Next, consider a diagonal element $(m,m)$:
  \begin{multline}
      0 =  \\
      \frac 12 z_{mm} + \frac 12 \sum_{j\neq m} z_{mj} + \frac 12 \sum_{j\neq m} z_{jm} + \frac 12 \sum_{j\neq m} z_{jm} + \frac 12 \sum_{j\neq m} z_{mj} \\
      = \frac 12 z_{mm},
   \end{multline}
  because all elements in the sums are zero.
  Hence all coefficients $z_{kj}$ must be zero. This proves linear independence.

\section{Parameterizing density matrices using $B^{kj}$.}\label{app:N2basismats}

{
The set of density matrices $\{B^{kj}\in \C^{N\times N}\}_{k,j=0,\dots,N-1}$ defined in \eqref{eq:basismats} form a basis of all Hermitian matrices in $\C^{N\times N}$ (see Appendix \ref{app:basismats_proof}).
Since the set of all density matrices is a convex subset of all Hermitian matrices~\cite{harriman1978geometry, bengtsson2013geometry}, any density matrix in $\C^{N\times N}$ can be written as a linear combination of those basis density matrices, with coefficients that sum up to one. 
}
{Therefore, the basis density matrices $B^{kj}$ define a parameterization of the set of all $N\times N$ density matrices, $\mathcal{D}_N$, where each element is a density matrix itself. The parameterization maps} coefficients $\boldsymbol{z}\in Q_N \subset \R^{N^2-1}$ to density matrices in $\mathcal{D}_N$:
\begin{align}
    F_N(\mathbf{z}) &= \sum_{k,j=0}^{N-1} z_{kj} B^{kj},
\end{align}
where {$z_{NN}$ is chosen such that $\sum_{kj}z_{kj} = 1$}. The set of admissible parameters $Q_N$ that yield a valid density matrix in $\mathcal{D}_N$ consists of all $\boldsymbol{z}$ in such a way that all the roots of the characteristic polynomial $\det(\rho - \lambda I)=0$ are non-negative, where $\rho =F_N(\boldsymbol{z})$.
The map $F_N\colon Q_N \to \mathcal{D}_N$ is bijective (one-to-one and onto), because the basis elements $B^{kj}$ span all Hermitian matrices.

Obviously, choosing non-negative coefficients, $z_{kj}\geq 0$ for $k,j=0,\dots,N-1$ that sum to one, always yields a valid density matrix in $\C^{N\times N}$. However, non-negativity of the coefficients is not necessary. {For example, for $N=2$, the coefficients $z_{00}=z_{11}=z_{01}=\frac 12$ and $z_{10} = -\frac 12$ yield a valid density matrix, even though $z_{10}<0$.}

For $N=2$, we can explicitly derive the set $Q_2$ by computing the eigenvalues of $F_2(\boldsymbol{z})$. The basis matrices $B^{kj}\in \C^{2\times 2}$ are given by 
\begin{align*}
    B^{00} &= \begin{bmatrix} 
                    1 & 0 \\ 0 & 0 
             \end{bmatrix}, \quad
             B^{01} = \frac 12 \begin{bmatrix} 
                    1 & 1 \\ 1 & 1 
             \end{bmatrix}, \\
    B^{10} &= \frac 12 \begin{bmatrix} 
                    1 & i \\ -i & 1 
             \end{bmatrix}, \quad
             B^{11} =  \begin{bmatrix} 
                    0 & 0 \\ 0 & 1 
             \end{bmatrix}. 
\end{align*}
Any density matrix $\rho\in \mathbb{C}^{2\times 2}$ is Hermitian, and can therefore be written in this basis with coefficients $z_{kj}\in \R$, where $k,j=0,1$:
\begin{multline}\label{eq_rho-sum}
    \rho = \sum_{k=0}^1\sum_{j=0}^1 z_{kj} B^{kj}\\
         = \begin{pmatrix}
            z_{00} + \frac 12 \left(z_{01} + z_{10}\right) & \frac 12 \left(z_{01} + i z_{10}\right) \\
            \frac 12 \left(z_{01} - i z_{10}\right)  & z_{11} + \frac 12 \left(z_{01} + z_{10}\right)
         \end{pmatrix}.
\end{multline}
Since $\Tr(\rho)=1$, we know that $\sum_{k,j} z_{kj}=1$. Further, $\rho$ is positive semi-definite, such that $Q_2$ can be derived from the eigenvalues $\lambda_{1,2}$ of $\rho$:
\begin{align*}
   0 &= \det\left(\rho - \lambda I_2\right)  \\
     &=\lambda^2  - \lambda \left(\rho_{00}+\rho_{11}\right)  + \rho_{00}\rho_{11} - \rho_{01}\rho_{10} \\
     \Rightarrow \lambda_{1,2} &= \frac{\rho_{00} + \rho_{11}}{2} \pm \sqrt{\left(\frac{\rho_{00} + \rho_{11}}{2}\right)^2 - \rho_{00}\rho_{11} + \rho_{01}\rho_{10}}.
\end{align*}
We first note that $\frac 12 \left(\rho_{00} + \rho_{11}\right) = \frac 12 \sum_{k,j}z_{kj} = \frac 12\geq 0$. The term under the square root is non-negative because $\lambda_{1,2}\in\R$ since $\rho^\dag = \rho$. Therefore, both eigenvalues $\lambda_{1,2}$ are non-negative if and only if
\begin{align*}
&\frac 12 \geq  \sqrt{\frac 14 - \rho_{00}\rho_{11} + \rho_{01}\rho_{10}}\\
\Leftrightarrow \quad &\rho_{00}\rho_{11} \geq \rho_{01}\rho_{10} \\
     \Leftrightarrow \quad & \left(z_{00} + \frac {z_{01} + z_{10}}{2} \right)  \left(z_{11} + \frac {z_{01} + z_{10}}{2} \right) \geq \frac {z_{01}^2 + z_{10}^2}{4} \\
     \Leftrightarrow \quad &z_{01}z_{10} \geq z_{00}^2 - z_{00} + z_{11}^2- z_{11},
\end{align*}
where the last line follows from $\frac 14 \left(z_{01}^2 + z_{10}^2\right) = \frac 14 \left(z_{01}+z_{10}\right)^2 - \frac 12 z_{01}z_{10}$ and $z_{01} + z_{10} = 1 - \left(z_{00} + z_{11}\right)$, and some basic algebra.
Substituting $z_{01} = 1 - z_{00} - z_{11}-z_{10}$ shows that $\lambda_{1,2} \geq 0$ if and only if
\begin{align}\label{eq:nonneglambda}
    z_{00}^2 + z_{11}^2 + 
    \left(z_{10} - 1 \right)\left(z_{00} + z_{11} + z_{10}\right) \leq 0
\end{align}
Necessary and sufficient conditions for the sum in \eqref{eq_rho-sum} to represent a density matrix are therefore: 1) the coefficients $z_{00}, z_{11}, z_{10}\in \R$ satisfy the inequality \eqref{eq:nonneglambda}, and 2) $z_{01} = 1 - z_{00} - z_{11}-z_{10}$.

\begin{figure}
    \centering
    \includegraphics[width=.48\textwidth]{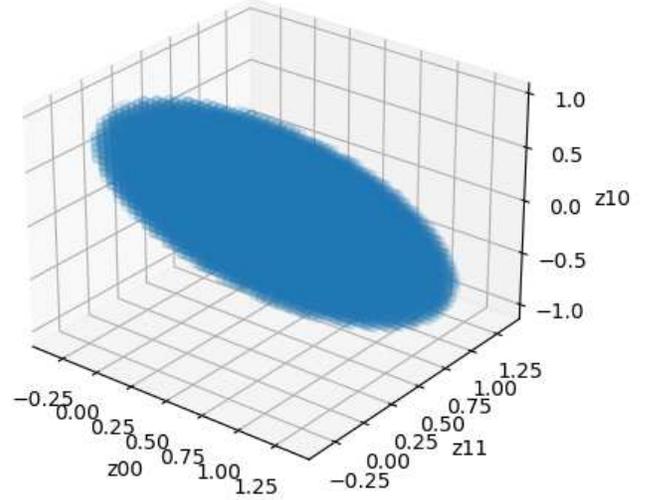}
    \caption{Set of parameters, $Q_2$, that generate all density matrices in $C^{2\times 2}$.}
    \label{fig:N2_coeffs}
\end{figure}

By introducing the rotated and translated coordinates
\begin{align*}
    \xi = z_{00} + \frac{(z_{10}-1)}{2},\quad
    \eta = z_{11} + \frac{(z_{10}-1)}{2},
\end{align*}
we can write the inequality \eqref{eq:nonneglambda} in the form
\begin{align*}
    2\xi^2 + 2\eta^2 + z_{10}^2 \leq 1.
\end{align*}
This is an ellipsoid with semi-axes $(1/\sqrt{2},\, 1/\sqrt{2},\, 1)$ in $(\xi, \, \eta,\, z_{10})$ coordinates.
The set of coefficients that parameterize any density matrix in $\C^{N\times N}$ for $N=2$ is therefore given by the three-dimensional ($(N^2-1)$-dimensional) set
\begin{eqnarray}\label{eq_convex-set}
    Q_2 &:= &\left\{\mathbf{z}=\begin{pmatrix} z_{00}\\  z_{11}\\ z_{10}\end{pmatrix} \in \R^3 \quad \text{such that} \right.\\
      &&\left. 2\left( z_{00}+\frac{z_{10}-1}{2}\right)^2  + 2\left( z_{11}+\frac{z_{10}-1}{2}\right)^2 +z_{10}^2 \leq 1 \right\}. \nonumber
\end{eqnarray}
The above expression yields a one-to-one and onto mapping $F_2 \colon Q_2 \to \mathcal{D}_2$ from coefficients $\mathbf{z}\in Q_2$ to the set $\mathcal{D}_2$ of all $2\times 2$ density matrices:
\begin{align}
    F_2(\mathbf{z}) &= \sum_{k,j=0}^{1} z_{kj} B^{kj}
\end{align}
where $z_{01} = 1 - z_{00} - z_{11}-z_{10}$. 
The set of admissible coefficients $Q_2$ is visualized in Figure~\ref{fig:N2_coeffs}. 

It would be desirable to extend the explicit derivation of $Q_N$ to higher dimensional cases with $N\geq 3$. However, deriving conditions to guarantee non-negative eigenvalues is non-trivial for $N\geq 3$ as it involves solving for the roots of the characteristic polynomial of degree $N$. 

\section{General pure-state preparation.}
\label{app:general_pure_state}

{
Section \ref{sec:optim} presents a numerical optimization framework for preparing pure target states that are represented by density matrices of the form $\rho_t = \boldsymbol{e}_m\boldsymbol{e}_m^\dagger$, where $\boldsymbol{e}_m\in\R^N$ is the $m$-th unit vector. In this appendix, we generalize the framework to the preparation of any pure state.
}

{
Let the density matrix for the general pure target state be $\rho_t = \psi_t\psi_t^\dagger$, where $\psi_t \in \C^N$ is the corresponding state vector. By rotating the basis by a unitary transformation $U^\dagger$, it is always possible to map the coordinates of the state vector to a unit vector such that $\phi_t = U \psi_t = \boldsymbol{e}_m$, for some $m\in[0,N-1]$. In this basis, the density matrix for the target state becomes diagonal, $\sigma_t = U \rho_t U^\dagger = \boldsymbol{e}_m\boldsymbol{e}_m^\dagger$.} 

{For a pure target state of that form, we have constructed the objective function in \eqref{eq:objective_Jexpected} to utilize
a diagonal positive semi-definite matrix $N_m$ such that the $m^{th}$ element is zero and all other elements are positive. For an arbitrary density matrix $\rho$ of size $N\times N$, the transformed matrix $U\rho U^\dagger$ is also a density matrix, and the objective function for this state is non-negative: 
\begin{align}\label{eq:transform}
    0 \leq \Tr(N_m U\rho U^\dagger) = \Tr(U^\dagger N_m U \rho).
\end{align}
Further, equality occurs if and only if $U\rho U^\dagger = \eb_m\eb_m^\dagger$ and hence $\rho = \psi_t\psi_t^\dagger = \rho_t$.
The general pure-state optimization problem therefore uses the transformed observable $N_m^\prime := U^\dagger N_m U$ in the objective function:
\begin{align}
    \min\,  J(\rho_s(T)) = \Tr(N'_m \rho_s(T)). 
\end{align}
As before, $\rho_s(T)$ solves Lindblad's master equation with the initial condition $\rho_s(0)$, see Equation~\eqref{eq:basis_superposition}. The same reasoning as in Section \ref{sec:optim} then applies. In particular, if the optimized control pulses give $J(\rho_s(T))=0$, any arbitrary initial state $\rho_i(0)=\sum_{kj}z_{kj}B^{kj}$ will also result in $\rho_i(T) = U^\dagger e_m e_m^\dagger U = \psi_t \psi_t^\dagger$.
}

\section{Driving a qudit-cavity system to the $|10\rangle$ state.}\label{app:results_purestateprep}

{
Here, we exemplify the application of the proposed framework to a pure-state preparation problem that is not the ground state. We aim to  drive a qudit from any initial state to its first excited state, while leaving the cavity in the ground state (both initially and at the final time). The system parameters are the same as in the first test case in Section \ref{sec:numerical_results}, given in Table \ref{tab:3x20_coeffs}. The initial condition $\rho_s(0)$ used throughout the optimization spans the basis in the qudit's dimension and couples to the ground state in the cavity, as given in Equation \eqref{eq:3x20_initcond}. In contrast to the ground-state optimization problem, the objective function now targets the pure state $|10\rangle$ of the coupled system, which corresponds to the unit vector $\boldsymbol{e}_{20}$ in $\R^{60}$ (i.e. $m=20$ in Eq.~\eqref{eq:objective_Jexpected}):
\begin{align}
    J(\rho_s(T)) = \Tr\left(N_{20}\rho_s(T)\right)
\end{align}
}

\begin{figure}
    \centering
    \input{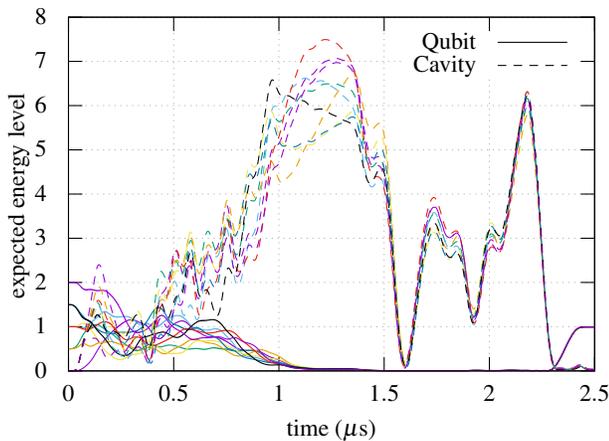}
    \caption{{Time-evolution of expected energy levels for various initial conditions using the optimized controls to drive the qudit-cavity system to the $|10\rangle$ state.}}
    \label{fig:3x20_purestate_expected}
\end{figure}

{
Figure \ref{fig:3x20_purestate_expected} plots the optimized time-evolution of the expected energy levels for the qudit and the cavity for various initial qudit states (the cavity starts in the ground state in this case). At final time $T=2.5\mu$s, the average fidelity of reaching the $|1\rangle$-state in the qudit is $99.43\%$, and for reaching the $|0\rangle$-state in the cavity is $99.10\%$.
}

\section*{References}
\bibliography{main.bib}

\end{document}